\documentclass[useAMS,usenatbib]{mn2e}
\usepackage{rotating}
\usepackage{longtable}
\usepackage{graphics}
\usepackage{times}

\def\teff{$T_{\rm eff}$}
\def\hd{HD\,177765}
\def\lgg{$\log g$}
\def\ha{H$\alpha$}
\def\ms{m\,s$^{-1}$}
\def\kms{km\,s$^{-1}$}
\def\i{\,{\sc i}}
\def\ii{\,{\sc ii}}
\def\iii{\,{\sc iii}}
\def\vsini{$v_{\rm e}\sin i$}
\def\bCrB{$\beta$ CrB}

\def\templogg{{\sc TempLogG$^{\rm TNG}$}}
\def\vald{{\sc VALD}}
\def\WIDTH{{\sc WIDTH9}}
\def\WIDTHmf{{\sc WIDTHmf}}
\def\dream{{\sc DREAM}}

\def\synthmag{{\sc SynthMag}}
\def\llm{{\sc LLmodels}}

\def\apj{ApJ}%
\def\apjl{ApJ}%
\def\apss{Ap\&SS}%
\def\aap{A\&A}%
\def\aaps{A\&AS}%
\def\mnras{MNRAS}%
\newcommand{\bs}{$\langle B \rangle$}

\newcommand{\fifps}[2]{\centering\resizebox{#1}{!}{\includegraphics{#2}}}

\newcommand{\mybf}[1]{{#1}}

\begin{document}

\title[The longest-period roAp star HD\,177765]
{Discovery of the longest-period rapidly oscillating Ap star HD\,177765%
\thanks{Based on observations collected at the European Southern Observatory, Paranal, Chile (program 085.D-0124).}
}

\author[D.~Alentiev et al.]
{D.~Alentiev$^{1,2}$,
O.~Kochukhov$^3$,
T.~Ryabchikova$^4$
M.~Cunha$^2$,
V.~Tsymbal$^1$,
W.~Weiss$^5$ \\
$^1$ Department of Physics, Tavrian National University, Vernadskiy's Avenue 4, 95007 Simferopol, Ukraine\\
$^2$ Centro de Astrofisica da Universidade do Porto, Rua das Estrelas, 4150-762 Porto, Portugal\\
$^3$ Department of Physics and Astronomy, Uppsala University Box 516, 751 20 Uppsala, Sweden\\
$^4$ Institute of Astronomy, Russian Academy of Sciences, Pyatnitskaya 48, 119017 Moscow, Russia \\
$^5$ Department of Astronomy, University of Vienna, T\"urkenschanzstrasse 17, 1180 Wien, Austria}

\date{Accepted . Received ; in original form }

\pagerange{\pageref{firstpage}--\pageref{lastpage}}
\pubyear{2011}

\maketitle

\label{firstpage}

\begin{abstract}
We present the discovery of a long-period, rapidly oscillating Ap star, \hd.
Using high-resolution time-series observations obtained with UVES at the ESO VLT telescope, we found radial velocity variations with amplitudes 7--150~\ms\ and a period of 23.6~min, exceeding that of any previously known roAp star. The largest pulsation amplitudes are observed for Eu\iii, Ce\iii\ and for the narrow core of H$\alpha$. We derived the atmospheric parameters and chemical composition of HD\,177765, showing this star to be similar to other long-period roAp stars. Comparison with theoretical pulsational models \mybf{indicates} an advanced evolutionary state for HD\,177765. \mybf{Abundance analyses of this and other roAp stars suggest} a systematic variation with age of the rare-earth line anomalies seen in cool Ap stars.
\end{abstract}


\section{Introduction}
\label{intro}

The rapidly oscillating (roAp) stars are magnetic, chemically peculiar stars which pulsate in high-overtone acoustic  modes with typical periods of $\approx$\,10~min. These stars are located close to the instability strip crossing the main sequence between the early F and late A spectral types. First roAp pulsators were discovered by \citet{1982MNRAS.200..807K}. Currently, about 40 of such stars are known.

Several excitation mechanisms were suggested in the past to drive pulsations in roAp stars  \citep{1988IAUS..123..291D,1984trss.conf..346D,1983ApJ...275L...5S,1988MNRAS.235P...7M}.  Currently, it is widely accepted that the high frequency oscillations observed in these stars are excited by the opacity mechanism working on the hydrogen ionization region.  However, full non-adiabatic calculations show that the excitation of  high frequency acoustic oscillations by this mechanism can only be achieved in non-standard models, such as models with a modified T-tau relation \citep{1998MNRAS.301...31G}, or models with envelope convection partially  or fully suppressed \citep{2001MNRAS.323..362B,2005MNRAS.360.1022S}.Among these, models with convection suppressed seem to reproduce better the observed instability strip, however, the predicted red edge remains significantly hotter than the observed one.

The majority of roAp stars have been identified and analysed using high-speed photometric methods \citep{2000BaltA...9..253K}. However, it has been realised that the ground-based photometry is not particularly suitable for discovering roAp stars. Time-resolved spectroscopy has a major advantage for detection of smaller-amplitude and longer-period pulsations. This has been demonstrated by, for example, the discovery of very low-amplitude pulsations in HD\,75445 \citep{2009A&A...493L..45K} and by the detection of long-period oscillations in HD\,137909 \citep[\bCrB,][]{2004MNRAS.351..663H} and HD\,116114  \citep{2005MNRAS.358..665E}, which have been repeatedly classified as non-pulsating by photometric observations \citep{1994MNRAS.271..129M,2005IBVS.5651....1L}.

The two latter objects, along with a faint roAp star KIC\,10195926 found by the Kepler satellite \citep{2011MNRAS.414.2550K}, form an unusual sub-group with pulsation periods of 16--21~min and spectroscopic properties different from the ``classical'', shorter-period roAp stars. Here we present the spectroscopic discovery of another member of this class, HD\,177765, whose pulsation period of 24~min is the longest known for any roAp star.

\section{Observations and data reduction}
\label{observ}

Time-resolved spectroscopic observations of \hd\ were carried out on 12 June 2010, using the Ultraviolet and Visual Echelle Spectrograph (UVES)
at one of the 8.2-m UTs of the ESO Very Large Telescope. We obtained 50 spectra during a 66-min observing period, which started at HJD=2455359.8087 and finished at HJD=2455359.8547. The spectrograph was used in the 600~nm red-arm setting with an image slicer. This configuration provided a resolving power of $R\approx110,000$ and complete wavelength coverage of the 4980--7010~\AA\ region with the exception of a 65~\AA-wide gap at $\lambda\approx5990$~\AA.

We used 60~s exposure times for individual observations. The employed ultra-fast (4-port, 625~kpix\,s$^{-1}$) readout mode of the UVES CCDs allowed to reduce the overhead to 21~s, giving a time resolution of 81~s. The typical signal-to-noise ratio (S/N) of individual spectra was 80-95 in the 5000--5500~\AA\ region. The stellar observations were followed by a single ThAr calibration exposure. In addition, bias and flat-field images were obtained as part of the standard daytime calibration of UVES.

Reduction of the spectra was perform with a new version of our UVES pipeline, described by \citet{2007smma.conf..100L} and used in our previous studies of roAp stars \citep{2007MNRAS.376..651K,2007A&A...473..907R}. This code takes care of the standard spectra processing and calibration operations: bias averaging and subtraction, determination of the echelle order positions, subtraction of the scattered light, extraction of 1-D stellar spectra, pixel-to-pixel sensitivity and blaze function correction using flat-field. Compared to the previous version of the code, we introduced a new procedure for the determination of the order boundaries and improved the calculation of the heliocentric Julian date and barycentric radial velocity correction.

In the last step, we determined the dispersion solution with a precision of 35--40~\ms\ using ThAr emission spectrum taken after the stellar
time-series and performed the final continuum normalization. The latter was carried out in two steps. First, we produced a high-quality average
spectrum and determined the continuum level by fitting a spline function to manually selected points. Second, we re-normalized automatically each
individual spectrum to match the continuum of the average observation.

\section{The properties of \hd}
\label{star}

Little was known about \hd\ before our study. This star, classified as A5 SrEuCr \citep{2009A&A...498..961R}, was observed by \citet{1994MNRAS.271..129M}, who did not detect pulsations from the photometric observations on two
nights. \citet{1996A&A...311..901M} estimated \teff\,=\,8060~K, while \citet{1997A&AS..123..353M} detected
magnetic field from the Zeeman split line Fe\ii\ $\lambda$ 6149.2~\AA. According to that study, the mean field modulus of \hd\ remained constant at \bs\,=\,3.4~kG with a scatter of 20~G during about 2 years covered by their observations, suggesting a very long rotational period.

We used \templogg\ code \citep{2006ASPC..349..257K} to determine atmospheric parameters of \hd\ from photometry. Using the Str\"{o}mgren photometric indices $(b-y)$\,=\,0.248, $m_1$\,=\,0.261, $c_1$\,=\,0.731, H$\beta$\,=\,2.834 reported by \citet{1993PhDT.......246M}, we obtained \teff\,=\,8050~K, \lgg\,=\,4.3
with the calibration by \citet{1985MNRAS.217..305M} and \teff\,=\,7900~K, \lgg\,=\,4.6 with the calibration by \citet{1993A&A...268..653N}. Simultaneously, we estimated the reddening $E(b-y)$\,=\,0.104 and metallicity [M/H]=+0.9. This interstellar reddening was applied to the Geneva photometric data\footnote{http://obswww.unige.ch/gcpd/ph13.html}. The dereddened $(B2-G)_{0}=-0.404$ index yields \teff\,=\,7750~K according to the calibration by \citet{2008A&A...491..545N}.

\begin{figure}
\centering
\fifps{7cm}{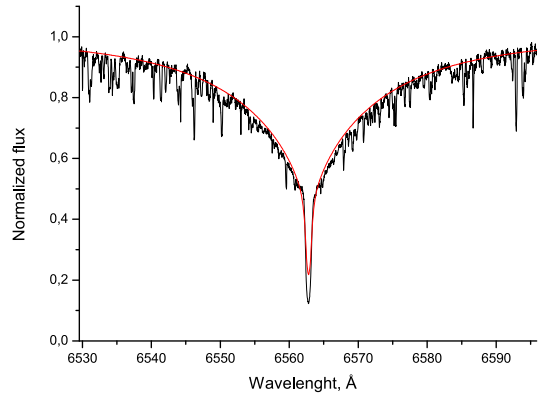}
\caption{Observed \ha\ profile (black line) and synthetic spectrum (red line) calculated for \teff\,=\,8000~K and \lgg\,=\,3.8.}
\label{Halpha}
\end{figure}

Taking into account the high metallicity estimated from the Str\"{o}mgren photometry and the similarity of the mean abundances of \hd\ and of the roAp star \bCrB\ (see Sect.\,\ref{Abund}), we calculated a set of models around \teff\,=\,8000~K and \lgg=3.5--4.0 with the \bCrB\ abundances using the \llm\ model atmosphere code \citep{2004A&A...428..993S}. The final atmospheric parameters of \hd\ were obtained by fitting the observed \ha\ profile to the synthetic spectra calculated using the {\sc SynthV} code \citep{1996ASPC..108..198T}. 
The best fit, corresponding to \teff\,=\,8000~K and \lgg\,=\,3.8, is illustrated in Fig.~\ref{Halpha}.

We estimated the mean magnetic field modulus \bs\,=\,3550~G from the separation of the Fe\ii\ $\lambda$ 6149.26~\AA\ Zeeman components in the average UVES spectrum. This measurement is significantly higher than the value reported by \citet{1997A&AS..123..353M}. We also used magnetic spectrum synthesis code \synthmag\ \citep{2007pms..conf..109K} to model partially resolved lines of different chemical elements. This analysis showed that a different field orientation is needed to fit the spectral lines of Fe-peak and rare-earth (REE) elements. To reproduce the latter, the field lines must be oriented predominantly parallel to the stellar surface, while the former lines require a significant radial field contribution. \mybf{This difference may be related to inhomogeneous horizontal abundance distributions.}

Due to the combined effect of magnetic field and chemical stratification it is difficult to derive an accurate projected rotational velocity for \hd. The commonly used magnetically insensitive Fe\i\ $\lambda$~5434.52~\AA\ line is too strong and is broadened due to a vertical abundance stratification. We estimated \mybf{the total broadening to be equivalent to} \vsini\,=\,2.2--2.7~\kms\ from the spectrum synthesis fit to a much weaker magnetically insensitive Fe\ii\ $\lambda$~6586.7~\AA\ line and to the partially resolved Zeeman components of several other weak lines. \mybf{In roAp stars this broadening is not necessarily due to stellar rotation alone \citep{2001A&A...374..615K,2007A&A...473..907R}.}

\begin{figure*}
\centering
\fifps{5.3cm}{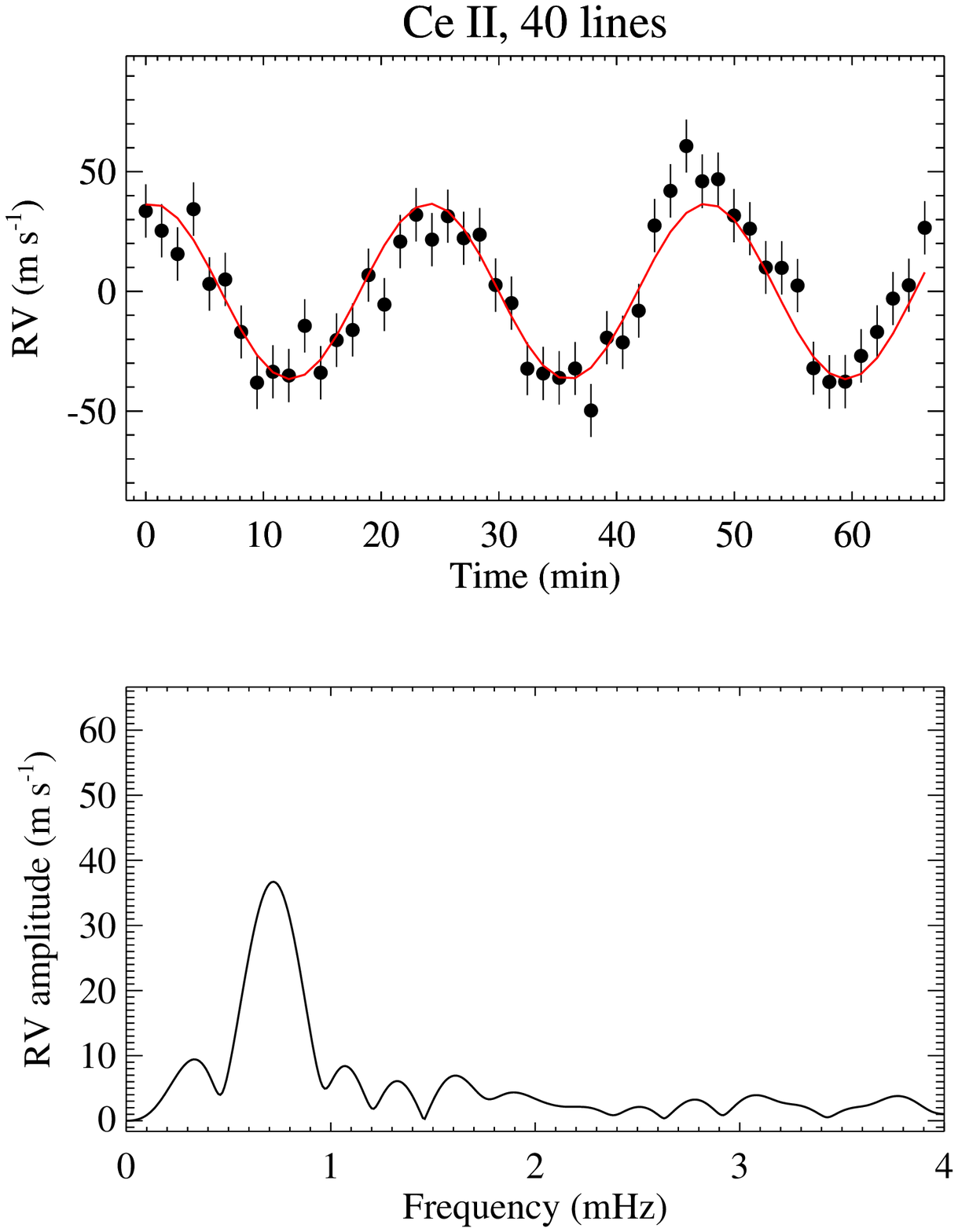}
\fifps{5.3cm}{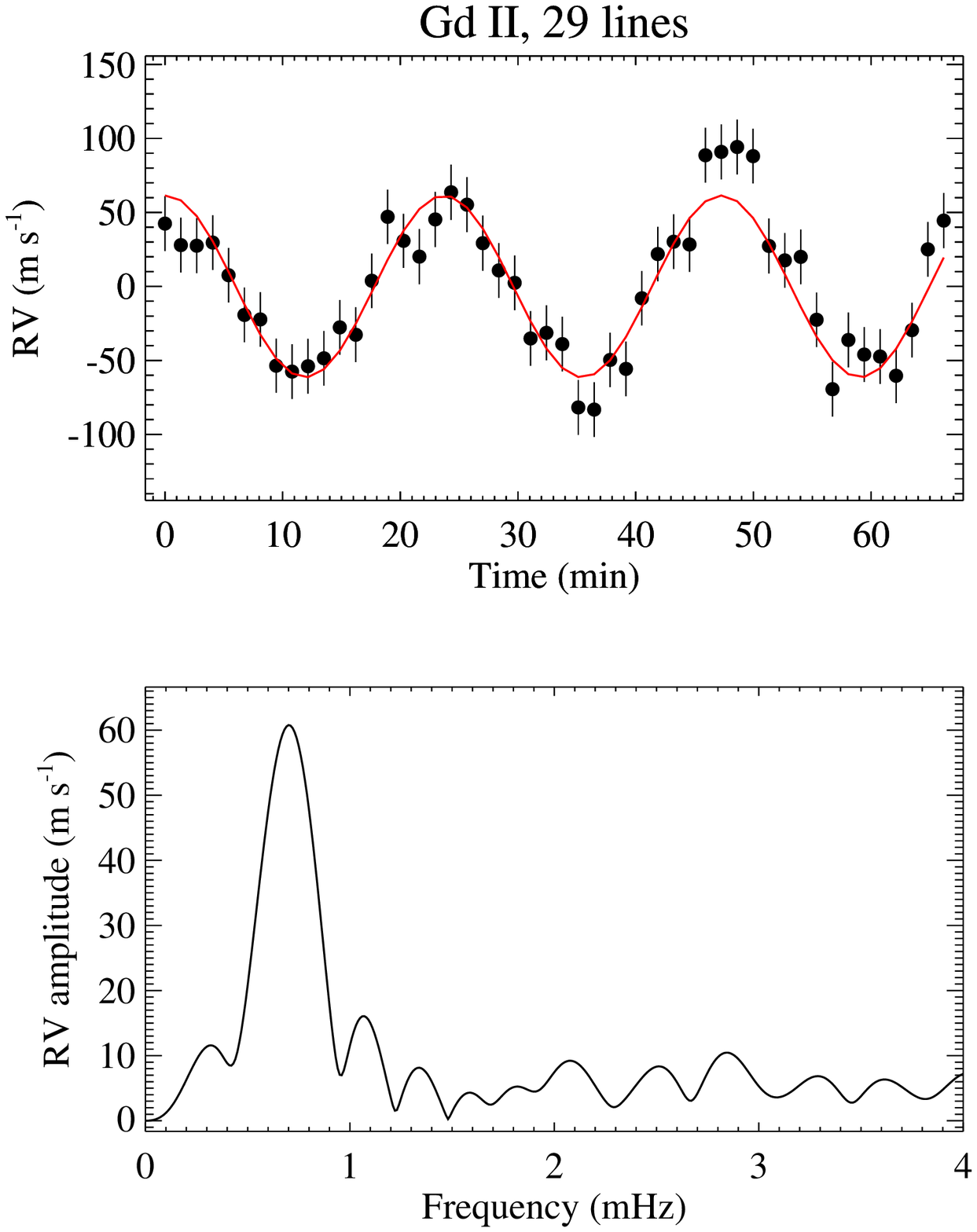}
\fifps{5.3cm}{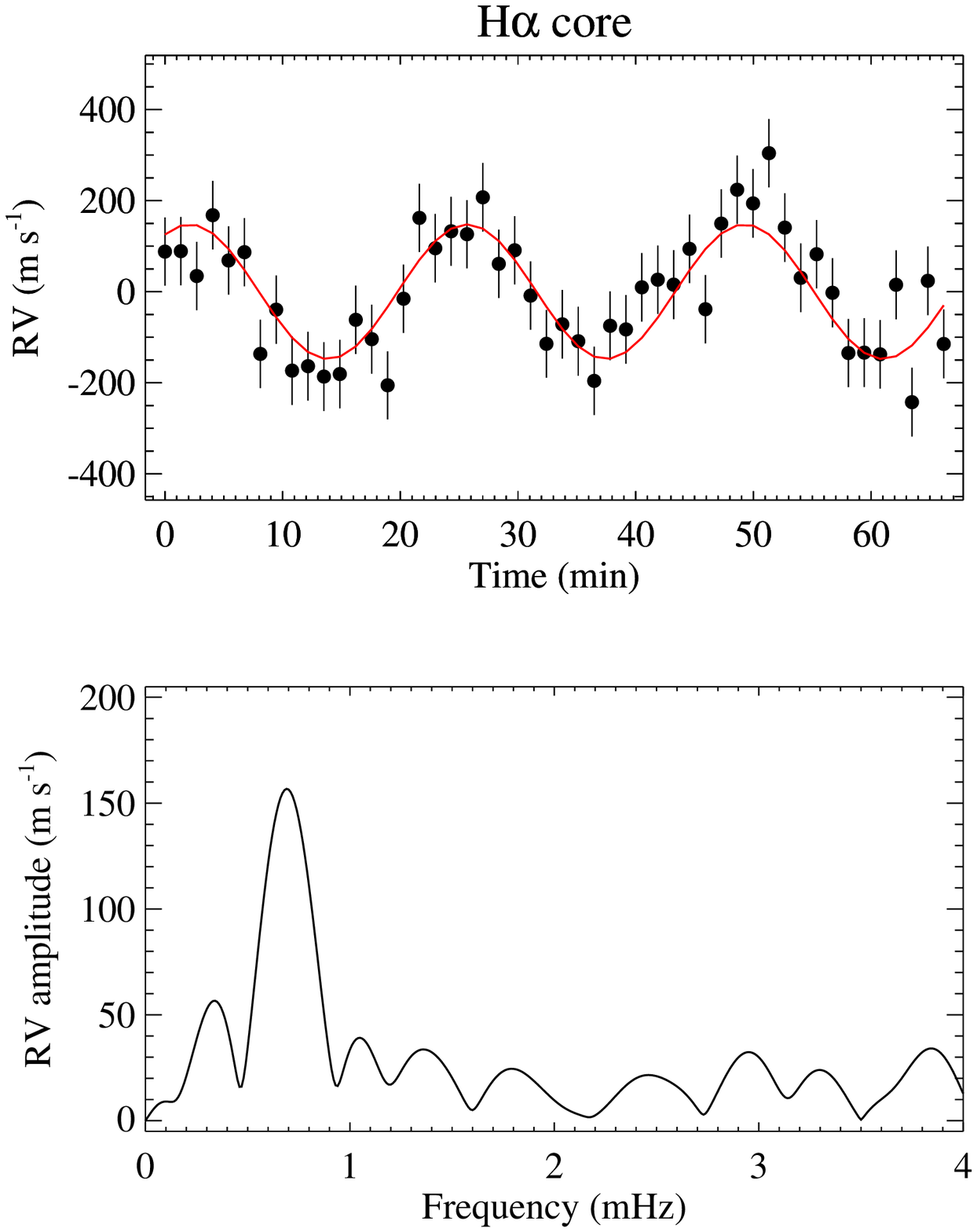}
\fifps{5.3cm}{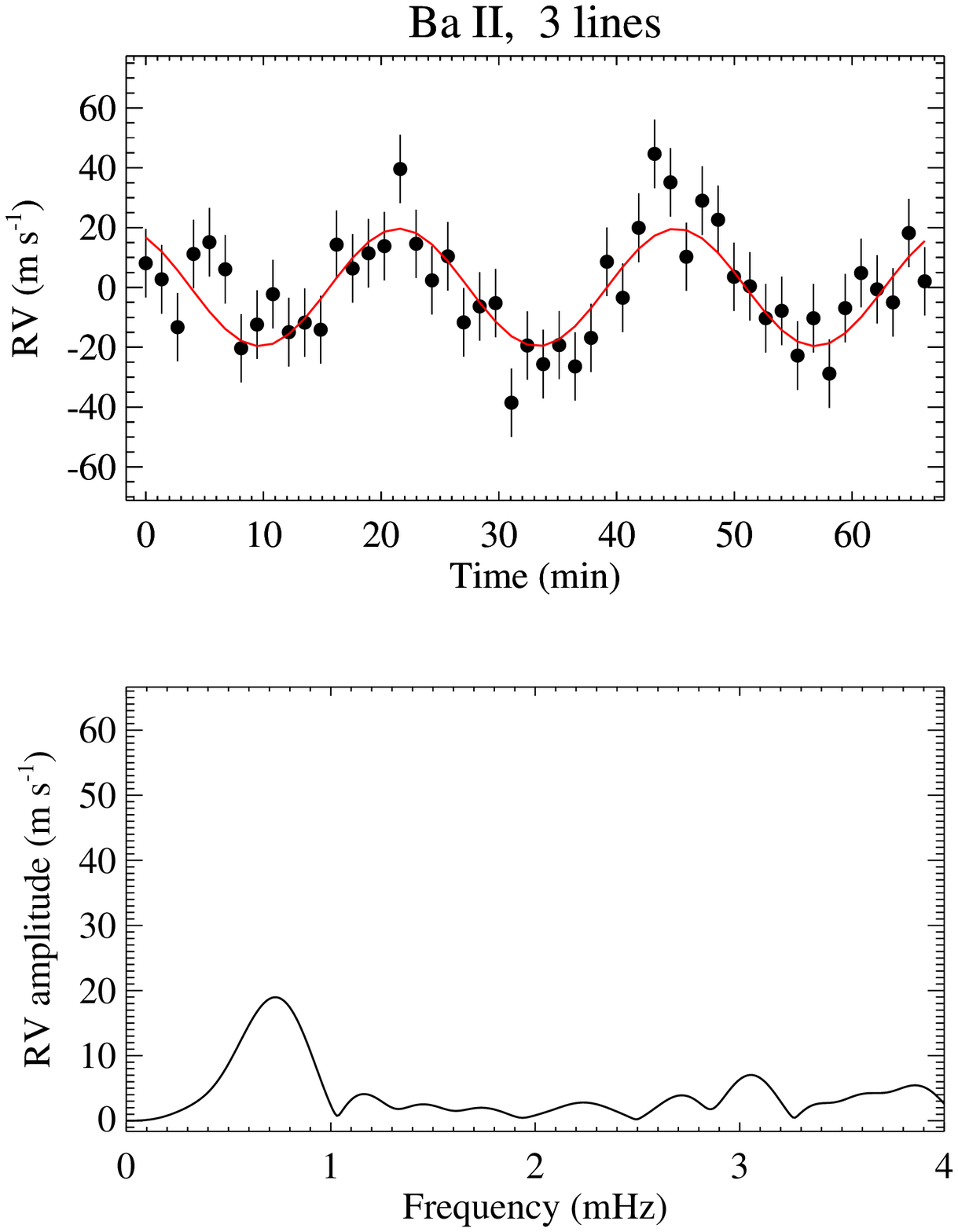}
\fifps{5.3cm}{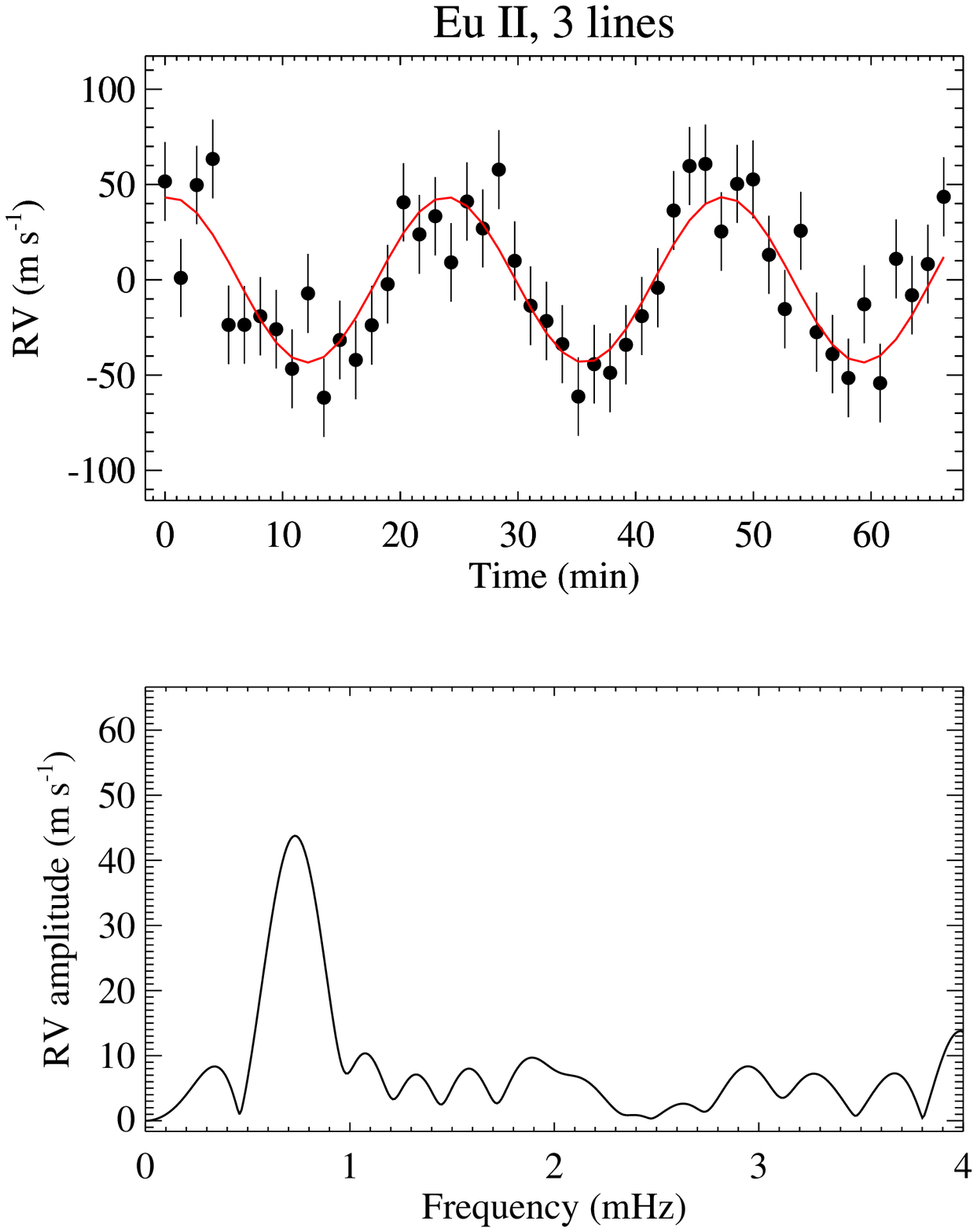}
\fifps{5.3cm}{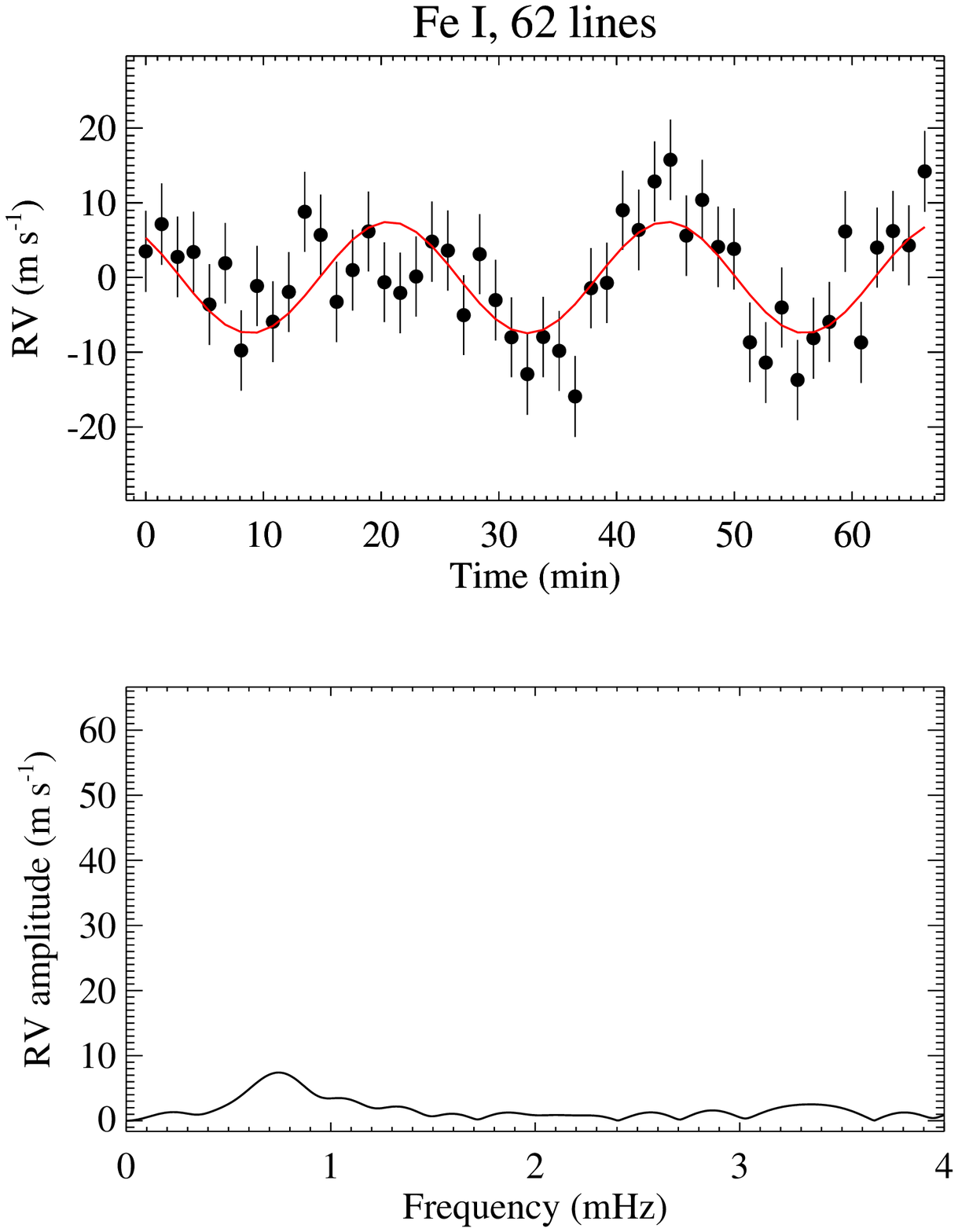}
\caption{Radial velocity variation and amplitude spectra for Ce\ii, Gd\ii, H$\alpha$ core, Ba\ii, Eu\ii, and Fe\i. Each panel shows the radial velocity curve on top, comparing the least-squares cosine fit (solid line) with observations (symbols). The corresponding amplitude spectra are presented below.}
\label{rv-freq}
\end{figure*}

\section{Radial velocity analysis}
\label{rv-analysis}

Initial analysis of the radial velocity (RV) data revealed variations in the core of H$\alpha$ with a period $>$\,20~min and suggested the presence of a marginal variability in many individual lines, especially those belonging to heavy elements. To obtain precise measurements, we combined RVs derived for all unblended lines of a given ion using the centre-of-gravity method \citep{2001A&A...374..615K}. Spectral lines were identified using information from the VALD data base \citep{1999A&AS..138..119K} and identification lists of variable lines compiled for other roAp stars \citep[e.g.,][]{2007A&A...462.1103R}.

Our frequency analysis consisted of the following steps. After obtaining mean RV measurements for each ion, we calculated the corresponding amplitude spectra using discrete Fourier transform and estimated an initial value for the pulsation period from the highest amplitude peak. We also computed a periodogram as described by \citet{1986ApJ...302..757H} in order to assess the False Alarm Probability (FAP) of the signal detection. Then we applied a non-linear least-squares fitting procedure to improve the period and estimate an amplitude and phase of the RV variations.

This analysis clearly showed the presence of pulsation variability (FAP\,$<$\,$10^{-5}$) in the core of H$\alpha$ and in the lines of Eu\ii, Gd\ii, and Ce\ii. The RV amplitude reaches 150~\ms\ for H$\alpha$, but it is only 40--60~\ms\ for the three rare-earth ions. These four elements show  periods of $22.85\pm0.39$~min (Eu), $23.50\pm0.26$~min (Ce), $23.85\pm0.27$~min (Gd), and $24.05\pm0.51$~min (H$\alpha$), yielding a weighted mean
pulsation period of $23.56\pm0.16$~min or a mean frequency of $\nu=0.707\pm0.005$~mHz, which is the lowest frequency detected in a roAp star. This 
period was adopted in the subsequent linear least-squares analysis of the remaining elements.

Several other ions show a probable ($10^{-5}<$\,FAP\,$<10^{-3}$) variation in the period range of 23--24~min with amplitudes of 10--130~\ms. A single line of Eu\iii\ shows the highest amplitude among metal lines. We also detected variability in the lines of Ba\ii, Yb\ii, and somewhat unexpectedly, Fe\i. Combining information from 62 lines of the neutral iron, we were able to detect the pulsation amplitude of $7.4\pm1.1$~\ms. At the same time, the RV curve constructed from 21 lines of ionized iron does not show any variation. Phase shifts of $\sim$\,0.1 of the pulsation period inferred from the RV curves of different ions probably reflect the difference in their formation heights.

Many roAp stars show large-amplitude pulsations in the lines of singly and doubly ionized Nd and Pr \citep{2007A&A...473..907R}. Nd\iii\ and Pr\iii\ lines are relatively weak and heavily blended in the spectrum of \hd, unlike in typical roAp stars where these lines are among the strongest metal spectral features \citep{2004A&A...423..705R}. Neither singly nor doubly ionized lines of Pr and Nd provide precise RVs for \hd\ due to
blending by iron peak elements. Our results indicate the absence of pulsation variability in the blends containing contributions by these ions with upper limits of $\approx$\,15--20~\ms. We also did not detect variability in the lines of Ca\i/\ii, La\ii, Cr\i/\ii, and Y\ii. A marginal signal at the right frequency may be present in the lines of neutral and ionized Ti.

Representative RV curves and amplitude spectra are shown in Fig.~\ref{rv-freq}. The outcome of the linear least-squares fit with a fixed pulsation period is reported in Table~\ref{ap-table} for all measured elements. This Table also gives the FAP information.

The short duration of our monitoring of \hd\ allowed to cover only 3 pulsation cycles. These observational data are insufficient to perform a very precise frequency analysis and assess possible presence of other frequencies. However, we note a systematic deviation of the mean RV curves of Ce\ii, Gd\ii, and H$\alpha$ core from the mono-periodic least-squares solution (see upper row in Fig.~\ref{rv-freq}). All three elements show a somewhat higher amplitude in the second half of the time-series, which indicates the presence of additional pulsation frequencies.

\begin{table}
\caption{Results of the time-series analysis of different ions in the spectrum of \hd. The columns give the element and ionization stage, the number of spectral lines used to construct average radial velocity curves, the amplitude $A$ and phase $\varphi$ derived in the least-squares cosine fit, where $\varphi$ is given as a fraction of the pulsation period $P=23.56$~min. The last column indicates a False Alarm Probability (FAP) for the highest peak in the corresponding amplitude spectra.
}
\label{ap-table}
\centering
\begin{tabular}{lcrccc}
\hline
\hline
Ion & N  & A (\ms) & $\varphi$ & FAP \\
\hline
Gd\ii       & 29 & 61.4$\pm$3.7~~   & 0.99$\pm$0.01 & 2.7e-8 \\
Ce\ii       & 40 & 36.7$\pm$2.2~~   & 0.98$\pm$0.01 & 2.6e-8 \\
H$\alpha$   & 1  &148.0$\pm$14.8    & 0.91$\pm$0.02 & 2.1e-7 \\
Eu\ii       & 3  & 43.3$\pm$4.1~~   & 0.99$\pm$0.02 & 8.7e-7 \\
Ba\ii       & 3  & 19.6$\pm$2.3~~   & 0.08$\pm$0.02 & 1.7e-5 \\
Yb\ii       & 3  & 31.3$\pm$4.3~~   & 0.01$\pm$0.02 & 1.2e-4 \\
Fe\i        & 62 &  7.4$\pm$1.1~~   & 0.12$\pm$0.02 & 1.7e-4 \\
Ce\iii      & 2  & 65.0$\pm$12.1    & 0.88$\pm$0.03 & 2.8e-4 \\
Ti\i+Ti\ii  & 12 &  8.8$\pm$1.7~~   & 0.06$\pm$0.03 & 5.3e-3 \\
Eu\iii      & 1  &128.8$\pm$28.5    & 0.06$\pm$0.03 & 3.0e-2 \\
Ca\i+Ca\ii  & 10 &  4.2$\pm$2.0~~   & 0.01$\pm$0.07 & 9.0e-2 \\
Fe\ii       & 21 &  3.8$\pm$1.7~~   & 0.10$\pm$0.07 & 2.5e-1 \\
La\ii       & 13 & 13.7$\pm$9.0~~   & 0.01$\pm$0.10 & 7.4e-1 \\
Cr\i+Cr\ii  & 12 &  7.1$\pm$2.2~~   & 0.19$\pm$0.05 & 3.4e-1 \\
Y\ii        & 9  & 11.0$\pm$5.1~~   & 0.30$\pm$0.08 & 6.3e-1 \\
Pr\iii      & 2  & 23.5$\pm$7.6~~   & 0.26$\pm$0.26 & 6.6e-1 \\
\hline
\end{tabular}
\end{table}

\begin{table}
\caption{Chemical composition of \hd. The error estimates are based on standard deviation of abundances estimated from N lines. The last column gives the abundances for \bCrB\ from \citet{2004A&A...423..705R}, except for Ce\iii\ and Eu\iii\ which were determined here.}
\label{abund-t}
\begin{center}
\begin{tabular}{l|ll|ll}
\hline
\hline
Ion &  $\log (N_{\rm el}/N_{\rm tot})$ & N & \bCrB  \\
\hline
C\i     & $-3.70$          & 1 &    \\
Si\i    & $-3.63\pm0.23$ & 3&\\
Si\ii   & $-3.56$          & 1& $-4.09$\\
Ca\i    & $-4.64\pm0.43$ & 6& $-5.10$\\
Ca\ii   & $-4.22$          & 1&\\
Ti\i    & $-5.94\pm0.24$ & 2& $-6.15$\\
Ti\ii   & $-6.48\pm0.19$ & 6& $-5.86$\\
Cr\i    & $-4.18\pm0.51$ &10& $-4.60$\\
Cr\ii   & $-4.36\pm0.35$ &24& $-4.68$\\
Mn\i    & $-5.07\pm0.04$ & 3&\\
Mn\ii   & $-5.30$     	 & 1& $-5.02$\\
Fe\i    & $-3.40\pm0.22$ &18& $-3.92$\\
Fe\ii   & $-3.25\pm0.32$ &23& $-3.66$\\
Co\i    & $-5.06\pm0.43$ & 4&        \\
Ni\i    & $-6.35\pm0.27$ & 3& $-5.41$\\
Sr\i    & $-5.47\pm0.05$ & 2& \\
Y\ii    & $-8.44$    	     & 1&\\
Zr\ii   & $-8.38$    	     & 1& $-8.39$\\
La\ii   & $-8.60\pm0.44$ & 3& $-8.35$\\
Ce\ii   & $-7.01\pm0.36$ &72& $-7.84$\\
Ce\iii  & $-5.49\pm0.09$ & 4 & $-5.65$\\
Pr\ii   & $-9.54$        & 1& $-9.26$\\
Pr\iii  & $-8.69$  	     & 1& $-9.35$\\
Nd\ii   & $-9.40\pm0.22$ & 2& $-9.17$\\
Nd\iii  & $-8.53\pm0.10$ & 3 & $-8.36$\\
Eu\ii   & $-7.95\pm0.13$ & 3& $-8.28$\\
Eu\iii  & $-6.20$          & 1& $-5.65$\\
Gd\ii   & $-7.62\pm0.29$ & 7& $-7.54$\\
Dy\ii   & $-6.90$          & 1&\\
Yb\ii   & $-7.99\pm0.31$ & 6&\\
\hline
\end{tabular}
\end{center}
\end{table}

\section{Abundance analysis}
\label{Abund}

We derived
preliminary abundance estimates for \hd\ from equivalent widths using a modified version of \WIDTH\ code (\WIDTHmf) written by V. Tsymbal,
where magnetic intensification effects are taken into account via the magnetic pseudo-microturbulence. We checked that this procedure works well for Fe, yielding a reasonable agreement with detailed magnetic spectrum synthesis calculations. For instance, fitting 18 Fe\i\ and 33 Fe\ii\ lines with \synthmag\ we obtained $\log (N_{\rm Fe}/N_{\rm tot})=-3.40\pm0.22$ for Fe\i\ and $\log (N_{\rm Fe}/N_{\rm tot})=-3.25\pm0.32$ for Fe\ii. The corresponding abundances retrieved with \WIDTHmf\ are $-3.65\pm0.33$ and $-3.40\pm0.40$. The systematic difference in abundances is well within the errors, which are rather large.

An inspection of individual Fe\ii\ lines reveals a strong dependence of the abundance on the line transition probability and on the excitation energy of the lower level.
We interpret this as an evidence for vertical chemical stratification, which is a common phenomenon for Ap stars. A detailed stratification analysis of \hd\ is beyond the scope of our study.

Atomic parameters for abundance determination were taken from \vald\ and from the \dream\ database for REEs \citep{1999Ap&SS.269..635B} using the \vald\ extraction tools. For Fe\ii\ lines a preference was given to the homogeneous set of calculations by \citet{1998A&A...340..300R}, supplemented by the data from \citet{2010A&A...520A..57C} for newly identified high excitation lines. The europium abundance was derived from Eu\ii\ lines by comparison with synthetic spectra taking magnetic field effects, isotopic and hyperfine splitting into account. The corresponding data were extracted from \citet{2001ApJ...563.1075L}. The parameters of the Eu\iii\ $\lambda$~6666.35 \AA\ line were adopted from \citet{2008A&A...483..339W}.

Our abundance estimates for \hd\ are given in Table\,\ref{abund-t}.
The last column of this table compares the chemical composition of \hd\ with the elemental abundances in the atmosphere of the roAp star \bCrB\ determined by \citet{2004A&A...423..705R}. These authors did not provide Ce\iii\ and Eu\iii\ abundances, therefore we have estimated them using the same model atmosphere and the same observations as in \citet{2004A&A...423..705R}. The similarity of the chemical composition of the two stars is clearly evident. Neither of these stars shows the 1.5--2.0 dex difference between abundances derived from Pr and Nd lines in the first and second ionization stages, which is typical of most roAp stars \citep{2004A&A...423..705R}. But they exhibit a pronounced CeEu ionization anomaly. Both \bCrB\ and \hd\ have longer pulsation periods than the stars with the PrNd anomaly.

\section{Conclusions and discussion}
\label{concl}

We have analysed time-resolved spectra of the cool Ap star \hd\ obtained with the UVES instrument at VLT. The radial velocity and frequency analysis of these data reveals this object to be a roAp star with the pulsation period of 23.6~min. These oscillations, clearly present with amplitudes 40--150~\ms\ in the combined radial velocity curves of Ce\ii, Eu\ii\ and Gd\ii\ lines, as well in the \ha\ core, occur with the longest known pulsational period for any roAp star. This discovery makes \hd\ a key object for testing predictions of pulsation theories because the frequency limits help in distinguishing alternative driving mechanisms and can provide useful asteroseismic constraints on the atmospheric and interior stellar structure.

No trigonometric parallax measurement is availble for \hd. Therefore, we can only approximately place it in the HR-diagram for a comparison with pulsational models. Using the effective temperature, derived from photometry and spectroscopy, the pulsation frequency $\nu=0.7$ mHz, and the evolutionary tracks from \citet{2002MNRAS.333...47C}, which provide the frequency of the most unstable mode for models with given effective temperature and luminosity, we obtain a mass $M$\,$\approx$\,2.2$M_\odot$ and  a luminosity $\log L/L_\odot\approx$\,1.5. These parameters imply that the star is significantly evolved from the zero age main sequence, making it similar to the three other long-period, evolved roAp stars: \bCrB, HD\,116114, and KIC\,10195926.

We have carried out an abundance analysis of \hd\ using the equivalent width and spectrum synthesis methods. 
A slow variation of the mean field modulus suggests a long rotational period. The diversity of Zeeman splitting patterns and the dependence of the abundance on the line strength and excitation potential (most clearly seen for Fe) indicates strong horizontal and vertical inhomogeneities.

We found that \hd\ is chemically similar to \bCrB. Both stars are distinguished by a high Ce abundance and discordant abundances inferred from different ions of \mybf{Eu and Ce}. But both stars lack a strong PrNd ionisation anomaly which is characteristic of higher frequency roAp stars. This anomaly is also absent in HD\,116114. Thus, instead of being a spectroscopic signature of roAp stars as suggested by \citet{2004A&A...423..705R}, the PrNd anomaly is probably related
to the evolutionary state of roAp stars, distinguishing evolved, long-period pulsators from the shorter-period stars closer to the zero age main sequence. This finding represents one of the most compelling evidence for a systematic variation of surface chemical composition of Ap stars with age.

Longer time-series observations of \hd\ are highly desirable to determine additional pulsation frequencies which are necessary for an astroseismic analysis. Observations during a single night might be sufficient to resolve the expected large frequency separation of $\Delta\nu_0\sim$\,50~$\mu$Hz.

\section*{Acknowledgments}
OK is a Royal Swedish Academy of Sciences Research Fellow, supported by grants from Knut and Alice Wallenberg Foundation and Swedish Research Council.
DA and MC acknowledge financial support of FCT/MCTES, Portugal, through the project PTDC/CTE-AST/098754/2008. MC is partially funded by POPH/FSE (EC).
TR acknowledges Presidium RAS Program ``Origin, structure and evolution of the objects in Universe'' for partial financial support.
WW was supported by the Austrian Science Fund (project P22691-N16).

\label{lastpage}

\end{document}